\documentclass[twocolumn,aps]{revtex4}

\usepackage{graphicx}     
\usepackage{dcolumn}      
\usepackage{bm}           

\begin{document}

\title{Implementation of Shor's Algorithm on a Linear Nearest Neighbour Qubit Array}

\author{Austin G. Fowler, Simon J. Devitt and Lloyd C. L. Hollenberg}

\affiliation{
Centre for Quantum Computer Technology, School of Physics\\
University of Melbourne, Victoria 3010, Australia.}
\date{\today}

\begin{abstract}
Shor's algorithm, which given appropriate hardware can factorise
an integer $N$ in a time polynomial in its binary length $L$, has
arguable spurred the race to build a practical quantum computer.
Several different quantum circuits implementing Shor's algorithm
have been designed, but each tacitly assumes that arbitrary pairs
of qubits within the computer can be interacted. While some
quantum computer architectures possess this property, many
promising proposals are best suited to realising a single line of
qubits with nearest neighbour interactions only. In light of this,
we present a circuit implementing Shor's factorisation algorithm
designed for such a linear nearest neighbour architecture. Despite
the interaction restrictions, the circuit requires just $2L+4$
qubits and to first order requires $8L^{4}$ gates arranged in a
circuit of depth $32L^{3}$ --- identical to first order to that
possible using an architecture that can interact arbitrary pairs
of qubits.
\end{abstract}


\maketitle

\section{Introduction}
\label{intro}

A quantum computer is a device that manipulates a collection of
small, interacting quantum systems. Usually each quantum system
contains just two states $|0\rangle$ and $|1\rangle$, and is
called a qubit. Unlike the bits in classical computers, qubits can
be placed in arbitrary superpositions $\alpha|0\rangle +
\beta|1\rangle$ and entangled with one another $(|00\rangle +
|11\rangle)/\sqrt{2}$. These two properties have enabled quantum
algorithms to be devised that are exponentially faster than their
classical equivalents \cite{Shor94a,Subr02}.

Building a practical device to implement quantum algorithms is a
daunting task. When devising a quantum algorithm it is frequently
assumed that any pair of qubits in the computer can be interacted.
However, many physical quantum computer proposals utilise short
range forces to couple qubits and hence only permit nearest
neighbour interactions
\cite{Wu99,Vrij00,Gold03,Nova03,Holl03,Tian03,Yang03,Feng03,Pach03,Frie03,Vand02,Soli03,Jeff02,Petr02,Golo02,Ladd02,Vyur00,Kame03}.
Indeed, each of the cited proposals recommends that just a single
line of qubits be built. Determining whether these linear nearest
neighbour (LNN) architectures can implement quantum algorithms in
a practical manner is a nontrivial and important question.

Implementing Shor's factorisation algorithm \cite{Shor94a} is
arguably the focus of much experimental quantum computer research
due to its encryption breaking applications. A necessary test of
any architecture is therefore whether or not it can implement
Shor's algorithm efficiently. In light of this, we present an LNN
circuit implementing Shor's algorithm requiring just $2L+4$ qubits
and $8L^{4}+40L^{3}+116\frac{1}{2}L^{2}+4\frac{1}{2}L-2$ gates
arranged in a circuit of depth $32L^{3}+80L^{2}-4L-2$.  The depth
of a circuit is the number of layers of gates in it, where a layer
of gates is a set of gates implemented in parallel.  The circuit
presented in this paper is based on the Beauregard circuit
\cite{Beau03}, which is designed for an architecture that can
interact arbitrary pairs of qubits.  To first order the Beauregard
circuit also has a gate count of $8L^{4}$ and circuit depth of
$32L^{3}$, provided one adds an additional qubit to the circuit to
allow repeated Toffoli gates to be implemented more quickly. The
precise differences are detailed throughout the paper.

The paper is structured as follows. In section \ref{shor_alg}
Shor's algorithm is briefly reviewed. In section \ref{shor_decomp}
Shor's algorithm is broken into a series of simple tasks. Section
\ref{can_dec} contains a brief description of the canonical
decomposition which is used to build fast 2-qubit gates. Sections
\ref{qft} to \ref{comp_circ} present, in order of increasing
complexity, the LNN quantum circuits that together comprise the
LNN Shor quantum circuit.  The LNN quantum Fourier transform (QFT)
is presented first, followed by a modular addition, the controlled
swap, modular multiplication, and finally the complete circuit.
Section \ref{conc} concludes with a summary of all results, and a
description of further work.

\section{Shor's Algorithm}
\label{shor_alg}

Shor's algorithm \cite{Shor94a} was published in 1994 and greeted
with great excitement due to its potential to break the popular
RSA encryption protocol \cite{Rive78}. RSA is used in all aspects
of e-commerce from Internet banking to secure online payment and
can also be used to facilitate secure message transmission. The
security of RSA is conditional on large integers being difficult
to factorise which has so far proven to be the case when using
classical computers. Shor's algorithm renders this problem
tractable.

To be specific, let $N = N_{1}N_{2}$ be a product of prime
numbers.  Let $L = \ln_{2}N$ be the binary length of $N$. Given
$N$, Shor's algorithm enables the determination of $N_{1}$ and
$N_{2}$ in a time polynomial in $L$.  This is achieved indirectly
by finding the period $r$ of $f(k) = m^{k} \bmod N$, where $m$ is
a randomly selected integer less than and coprime to $N$.  A
classical computer can then use $N$, $m$, and $r$ to determine
$N_{1}$ and $N_{2}$.

Quantum circuits implementing Shor's algorithm can be designed for
conceptual simplicity \cite{Vedr96}, speed \cite{Goss98}, minimum
number of qubits \cite{Beau03} or a tradeoff between speed and
number of qubits \cite{Zalk98}.  Table \ref{table:one} summarises
the various qubit counts and gate depths.  Note that generally
speaking time can be saved at the cost of more qubits.
\begin{table}
\begin{tabular}{c|c|c}
Circuit & Qubits & Depth \\
\hline
Simplicity \cite{Vedr96} & $\sim 5L$ & $O(L^{3})$ \\
Speed \cite{Goss98} & $O(L^{2})$ & $O(L \log L)$ \\
Qubits \cite{Beau03} & $\sim 2L$ & $\sim 32L^{3}$ \\
Tradeoff \cite{Zalk98}& $\sim 50L$ & $\sim 2^{17}L^{2}$
\end{tabular}
\caption{Number of qubits required and circuit depth of different
implementations of Shor's algorithm.  Where possible, figures are
accurate to first order in $L$.} \label{table:one}
\end{table}

An underlying procedure common to all implementations does exist.
The first common step involves initializing the quantum computer
to a single pure state $|0\rangle_{2L}|0\rangle_{L}$. Note that
for clarity the computer state has been broken into a $2L$ qubit
$k$ register and an $L$ qubit $f$ register.  The meaning of this
will become clearer below.  The various optimisations used to
achieve the qubit count of $2L+4$ discussed in this paper will be
presented in section \ref{comp_circ}.

Step two is to Hadamard transform each qubit in the $k$ register
yielding
\begin{equation}
\label{eq:two}
\frac{1}{2^{L}}\sum_{k=0}^{2^{2L}-1}|k\rangle_{2L}|0\rangle_{L}.
\end{equation}

Step three is to calculate and store the corresponding values of
$f(k)$ in the $f$ register
\begin{equation}
\label{eq:three}
\frac{1}{2^{L}}\sum_{k=0}^{2^{2L}-1}|k\rangle_{2L}|f(k)\rangle_{L}.
\end{equation}
Note that this step requires additional ancilla qubits. The exact
number depends heavily on the precise details of the
implementation.

Step four can actually be omitted but it explicitly shows the
origin of the period $r$ being sought. Measuring the $f$ register
yields
\begin{equation}
\label{eq:four} \frac{\sqrt{r}}{2^{L}}\sum_{n=0}^{
2^{2L}/r-1}|k_{0}+nr\rangle_{2L}|f_{0}\rangle_{L}
\end{equation}
where $f_{0}$ is the measured value and $k_{0}$ is the smallest
value of $k$ such that $f(k)=f_{0}$.

Step five is to apply the quantum Fourier transform
\begin{equation}
\label{eq:qft1} |k\rangle \rightarrow
\frac{1}{2^{L}}\sum_{j=0}^{2^{2L}-1}\exp(2\pi ijk/2^{2L})|j\rangle
\end{equation}
to the $k$ register resulting in
\begin{equation}
\label{eq:fivesum}
\frac{\sqrt{r}}{2^{2L}}\sum_{j=0}^{2^{2L}-1}\sum_{p=0}^{
2^{2L}/r-1}\exp(2\pi
ij(k_{0}+pr)/2^{2L})|j\rangle_{2L}|f_{0}\rangle_{L}.
\end{equation}
The probability of measuring a given value of $j$ is thus
\begin{equation}
\label{eq:prj} {\rm
Pr}(j,r,L)=\left|\frac{\sqrt{r}}{2^{2L}}\sum_{p=0}^{
2^{2L}/r-1}\exp(2\pi ijpr/2^{2L})\right|^{2}.
\end{equation}

If $r$ divides $2^{2L}$ Eq.~(\ref{eq:prj}) can be evaluated
exactly. The probability of observing $j=c2^{2L}/r$ for some
integer $0\leq c<r$ is $1/r$ whereas if $j\neq c2^{2L}/r$ the
probability is 0. When $r$ is not a power of 2 all one can say is
that with high probability the value $j$ measured will satisfy
$j\simeq c2^{2L}/r$ for some $0\leq c<r$. In either case, given a
measurement $j\simeq c2^{2L}/r$ with $c\neq 0$, information about
$r$ can be extracted via simple classical manipulations
\cite{Fowl03b}. Note that is quite likely that $r$ will not be
completely determined after just one run of the steps described
above and that further runs will be required.  Even when the final
value of $r$ is determined, if $r$ is not even or $r$ does not
satisfy $f(r/2)\neq N-1$ the entire process needs to be repeated
for a different value of $m$ in an attempt to get a different
value of $r$.  When a value of $r$ is found with the appropriate
properties, the factors of $N$ can be determined from $N_{1}={\rm
gcd}(f(r/2)+1,N)$ and $N_{2}={\rm gcd}(f(r/2)-1,N)$.

\section{Decomposing Shor's algorithm}
\label{shor_decomp}

The purpose of this section is to break Shor's algorithm into a
series of steps that can be easily implemented as quantum
circuits.  Neglecting the classical computations and optional
measurement step described in the previous section, Shor's
algorithm has already been broken into four steps.
\begin{enumerate}
\item Hadamard transform. \item Modular exponentiation. \item
Quantum Fourier transform. \item Measurement.
\end{enumerate}
The modular exponentiation step is the only one that requires
further decomposition.

The calculation of $f(k) = m^{k} \bmod N$ is firstly broken up
into a series of controlled modular multiplications.
\begin{equation}
f(k) = \prod_{i=0}^{2L-1}(m^{2^{i}k_{i}} \bmod N),
\label{eq:mult_series}
\end{equation}
where $k_{i}$ denotes the $i$th bit of $k$. If $k_{i}=1$ the
multiplication $m^{2^{i}} \bmod N$ occurs, and if $k_{i}=0$
nothing happens.

There are many different ways to implement controlled modular
multiplication \cite{Vedr96,Goss98,Zalk98,Beau03}.  The methods of
\cite{Beau03} require the fewest qubits and will be used in this
paper.  To illustrate how each controlled modular multiplication
proceeds, let $a(i)=m^{2^{i}} \bmod N$ and
\begin{equation}
x(i) = \prod_{j=0}^{i-1}(m^{2^{j}k_{j}} \bmod N).
\end{equation}
$x(i)$ represents a partially completed modular exponentiation and
$a(i)$ the next term to multiply by. Let $|x(i),0\rangle$ denote a
quantum register containing $x(i)$ and another of equal size
containing 0. Firstly, add $a(i)$ modularly multiplied by the
first register to second register if and only if (iff) $k_{i}=1$.
\begin{eqnarray}
|x(i),0\rangle & \mapsto & |x(i),0+a(i)x(i) \bmod N \rangle
\nonumber \\
& = & |x(i),x(i+1)\rangle.
\end{eqnarray}
Secondly, swap the registers iff $k_{i}=1$.
\begin{equation}
|x(i),x(i+1)\rangle\mapsto|x(i+1),x(i)\rangle
\end{equation}
Thirdly, subtract $a(i)^{-1}$ modularly multiplied by the first
register from the second register iff $k_{i}=1$.
\begin{eqnarray}
& & |x(i+1),x(i)\rangle \nonumber \\
& \mapsto & |x(i+1),x(i)-a(i)^{-1}x(i+1) \bmod
N\rangle \nonumber \\
& = & |x(i+1),0\rangle.
\end{eqnarray}
Note that while nothing happens if $k_{i}=0$, by the definition of
$x(i)$ the final state in this case will still be
$|x(i+1),0\rangle$.

The first and third steps described in the previous paragraph are
further broken up into series of controlled modular additions and
subtractions respectively.
\begin{eqnarray}
0+a(i)x(i) & = & 0+\sum_{j=0}^{L-1}a(i)2^{j}x(i)_{j} \bmod N, \label{eq:add_series_1} \\
\lefteqn{x(i)-a(i)^{-1}x(i+1) =} \hspace{18mm} \nonumber \\
\lefteqn{x(i)-\sum_{j=0}^{L-1}a(i)^{-1}2^{j}x(i+1)_{j} \bmod N,}
\hspace{10mm} \label{eq:add_series_2}
\end{eqnarray}
where $x(i)_{j}$ and $x(i+1)_{j}$ denote the $j$th bit of $x(i)$
and $x(i+1)$ respectively. Note that the additions associated with
a given $x(i)_{j}$ can only occur if $x(i)_{j}=1$ and similarly
for the subtractions. Given that these additions and subtractions
form a multiplication that is conditional on $k_{i}$, it is also
necessary that $k_{i}=1$.

Further decomposition will be left for subsequent sections.

\section{Canonical Decomposition}
\label{can_dec}

A crucial part of building efficient quantum circuits is building
efficient 2-qubit gates.  This is particularly true for LNN
circuits in which it is common to follow productive gates such as
controlled-NOT (CNOT) or controlled-phase (cphase) with swap gates
designed to bring other pairs of qubits together to be interacted.
Such pairs of 2-qubit gates can and should be combined into a
single compound gate.

The canonical decomposition enables any 2-qubit operator $U_{AB}$
to be expressed (non-uniquely) in the form $V_{A}^{\dag}\otimes
V_{B}^{\dag}U_{d}U_{A}\otimes U_{B}$ where $U_{A}$, $U_{B}$,
$V_{A}$ and $V_{B}$ are single-qubit unitaries and
$U_{d}=\exp[i(\alpha_{x}X\otimes X+\alpha_{y}Y\otimes
Y+\alpha_{z}Z\otimes Z)]$ \cite{Krau01}. Moreover, any entangling
interaction can be used to create an arbitrary $U_{d}$ up to
single-qubit rotations \cite{Brem02}.

\begin{figure}
\begin{center}
\resizebox{70mm}{!}{\includegraphics{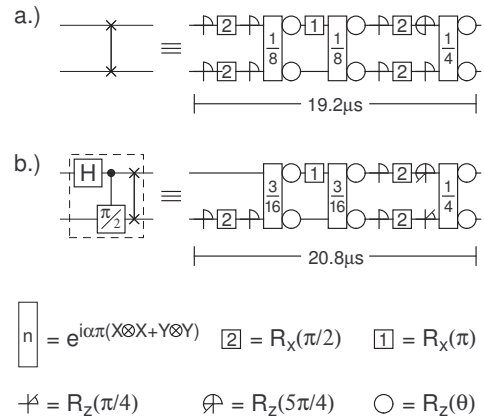}}
\end{center}
\caption{a.) Swap gate expressed as a sequence of physical
operations via the canonical decomposition. b.) Similarly
decomposed compound gate consisting of a Hadamard gate, controlled
phase rotation, and swap gate.  Note that the Kane architecture
\cite{Kane98} has been used for illustrative purposes.}
\label{figure:swap_hphases}
\end{figure}

Fig.~\ref{figure:swap_hphases}a shows the form of a swap gate
decomposed into a series of physical operations via the canonical
decomposition \cite{Hill03}. The Kane architecture \cite{Kane98}
has been used for illustrative purposes. Note that the full
circles in the figure represent Z-rotations of angle dependent on
the physical construction of the computer.
Fig.~\ref{figure:swap_hphases}b shows an implementation of the
composite gate Hadamard followed by a controlled $\pi/2$ phase
rotation followed by a swap. Note that the total time of the
compound gate is almost the same as the swap gate on its own.  In
certain cases, the total time of the compound gate can be much
less than the time required to implement any one of its
constituent gates \cite{Fowl03c}. In this paper, every sequence of
1- and 2-qubit gates that are applied to the same two qubits has
been implemented as a canonically decomposed compound gate.

\section{Quantum Fourier Transform}
\label{qft}

The first circuit that needs to be described, as it will be used
in all subsequent circuits, is the QFT.
\begin{equation}
\label{eq:qft2} |k\rangle \rightarrow
\frac{1}{\sqrt{2^{L}}}\sum_{j=0}^{2^{L}-1}\exp(2\pi
ijk/2^{L})|j\rangle
\end{equation}

\begin{figure}
\begin{center}
\resizebox{70mm}{!}{\includegraphics{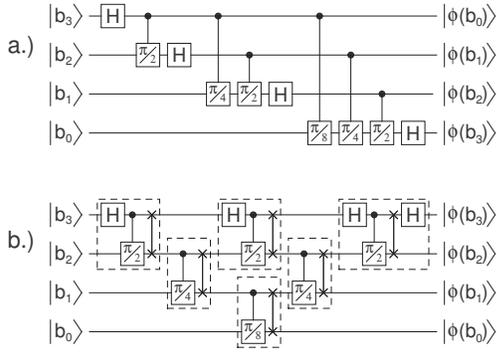}}
\end{center}
\caption{a.) Standard quantum Fourier transform circuit. b.) An
equivalent linear nearest neighbour circuit.}
\label{figure:serial_parallel_qft}
\end{figure}

Fig.~\ref{figure:serial_parallel_qft}a shows the usual circuit
design for an architecture that can interact arbitrary pairs of
qubits. Fig.~\ref{figure:serial_parallel_qft}b shows the same
circuit rearranged with the aid of swap gates to allow it to be
implemented on an LNN architecture. Dashed boxes indicate compound
gates.  Note that the general circuit inverts the most significant
to least significant ordering of the qubits whereas the LNN
circuit does not.

Counting compound gates as one, the total number of gates required
to implement a QFT on both the general and LNN architectures is
$L(L-1)/2$.  Assuming gates can be implemented in parallel, the
minimum circuit depth for both is $2L-3$.

\section{Modular Addition}
\label{mod_add}

\begin{figure}
\begin{center}
\resizebox{70mm}{!}{\includegraphics{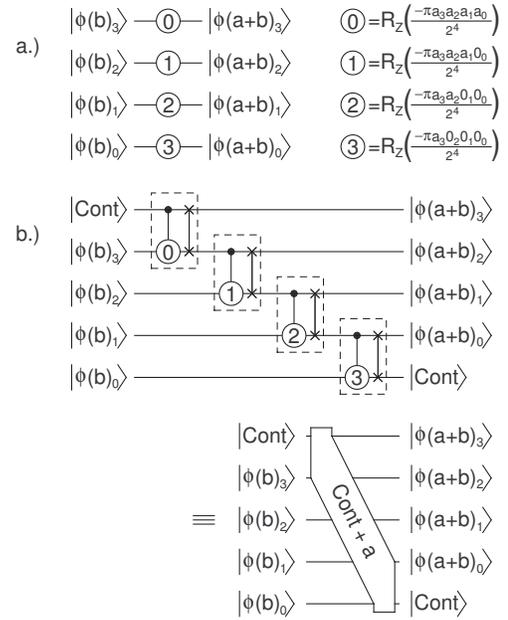}}
\end{center}
\caption{a.) Quantum Fourier addition. b.) Controlled quantum
Fourier addition and its symbolic equivalent circuit.}
\label{figure:fourier_add}
\end{figure}

Given a quantum register containing an arbitrary superposition of
binary numbers, there is a particularly easy way to add a binary
number to each number in the superposition \cite{Beau03}. By
quantum Fourier transforming the superposition, the addition can
be performed simply by applying appropriate single-qubit rotations
as shown in fig.~\ref{figure:fourier_add}a. Such an addition can
also very easily be made dependant on a single control qubit as
shown in fig.~\ref{figure:fourier_add}b.

\begin{figure*}[p]
\begin{center}
\resizebox{!}{215mm}{\includegraphics{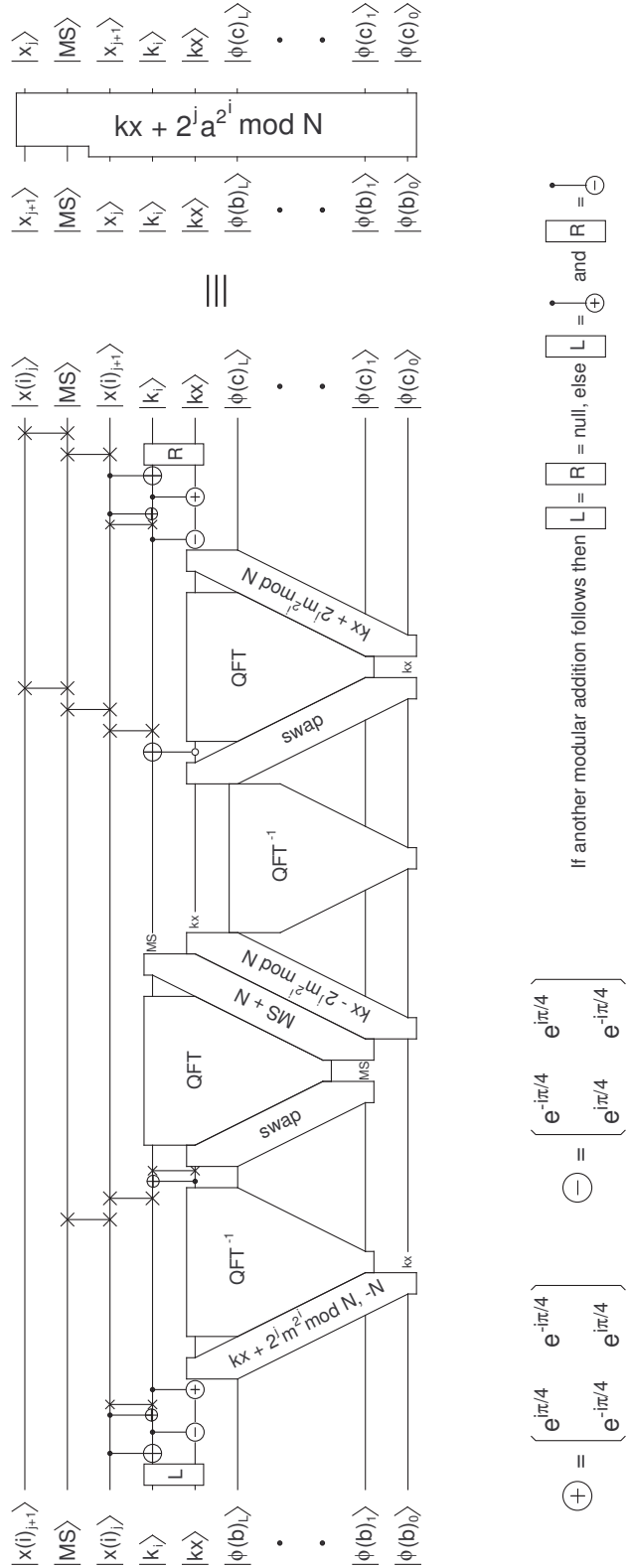}}
\end{center}
\caption{Circuit to compute $c = (b + 2^{j}m^{2^{i}}) \bmod N$.
The diagonal circuit elements labelled swap represent a series of
2-qubit swap gates.  Small gates spaced close together represent
compound gates.  The qubits $x(i)$ are defined in
eq.~\ref{eq:mult_series} and essentially store the current
partially calculated value of the modular exponentiation that
forms the heart of Shor's algorithm.  The $MS$ (Most Significant)
qubit is used to keep track of the sign of the partially
calculated modular addition result.  The $k_{i}$ qubit is the
$i$th bit of $k$ in eq.~\ref{eq:three}.  The $kx$ qubit is set to
1 if and only if $x(i)_{j} = k_{i} = 1$.} \label{figure:mod_add}
\end{figure*}

Performing controlled modular addition is considerably more
complicated as shown in fig.~\ref{figure:mod_add}. This circuit
adds $2^{j}m^{2^{i}} \bmod N$ to the register containing $\phi(b)$
to obtain $\phi(c)=\phi(b+2^{j}m^{2^{i}} \bmod N)$ iff both
$x(i)_{j}$ and $k_{i}$ are 1.  The first five gates comprise a
Toffoli gate that sets $kx=1$ iff $x(i)_{j}=k_{i}=1$.  $k_{i}$ and
$x(i)_{j}$ are defined in eq.~(\ref{eq:mult_series}) and
eqs.~(\ref{eq:add_series_1}-\ref{eq:add_series_2}) respectively.
Note that the Beauregard circuit does not have a $kx$ qubit, but
without it the singly-controlled Fourier additions become
doubly-controlled and take four times as long.  The calculations
of the gate count and circuit depth of the Beauregard circuit
presented in this paper have therefore been done with a $kx$ qubit
included.

The next circuit element firstly adds $2^{j}m^{2^{i}} \bmod N$ iff
$kx=1$ then subtracts $N$. If $b+(2^{j}m^{2^{i}} \bmod N) < N$,
subtracting $N$ will result in a negative number. In a binary
register, this means that the most significant bit will be 1.  The
next circuit element is an inverse QFT which takes the addition
result out of Fourier space and allows the most significant bit to
be accessed by the following CNOT. The $MS$ (Most Significant)
qubit will now be 1 iff the addition result was negative. If
$b+(2^{j}m^{2^{i}} \bmod N)> N$, subtracting $N$ will yield the
positive number $(b+2^{j}m^{2^{i}}) \bmod N$ and the $MS$ qubit
will remain set to 0.

We now encounter the first circuit element that would not be
present if interactions between arbitrary pairs of qubits were
possible.  Note that while this ``long swap'' operation
technically consists of $L$ regular swap gates, it only increases
the depth of the circuit by 1. The subsequent QFT enables the $MS$
controlled Fourier addition of $N$ yielding the positive number
$(b+2^{j}m^{2^{i}}) \bmod N$ if $MS=1$ and leaving the already
correct result unchanged if $MS=0$.

While it might appear that we now done, the qubits $MS$ and $kx$
must be reset so they can be reused. The next circuit element
subtracts $2^{j}m^{2^{i}} \bmod N$.  The result will be positive
and hence the most significant bit of the result equal to 0 iff
the very first addition $b+(2^{j}m^{2^{i}} \bmod N)$ gave a number
less than $N$.  This corresponds to the $MS=1$ case.  After
another inverse QFT to allow the most significant bit of the
result to be accessed, the $MS$ qubit is reset by a CNOT gate that
flips the target qubit iff the control qubit is 0.  Note that the
long swap operation that occurs in the middle of all this to move
the $kx$ qubit to a more convenient location only increases the
depth of the circuit by 1.

After adding back $2^{j}m^{2^{i}} \bmod N$, the next few gates
form a Toffoli gate that resets $kx$.  The final two swap gates
move $x(i)_{j+1}$ into position ready for the next modular
addition.  Note that the $L$ and $R$ gates are inverses of one
another and hence not required if modular additions precede and
follow the circuit shown.  Only one of the final two swap gates
contributes to the overall depth of the circuit.

The total gate count of the LNN modular addition circuit is
$2L^{2}+8L+22$ and compares very favourably with the general
architecture gate count of $2L^{2}+6L+14$. Similarly, the LNN
depth is $8L+16$ versus the general depth of $8L+13$.

\section{Controlled swap}
\label{cswap}

\begin{figure}
\begin{center}
\resizebox{70mm}{!}{\includegraphics{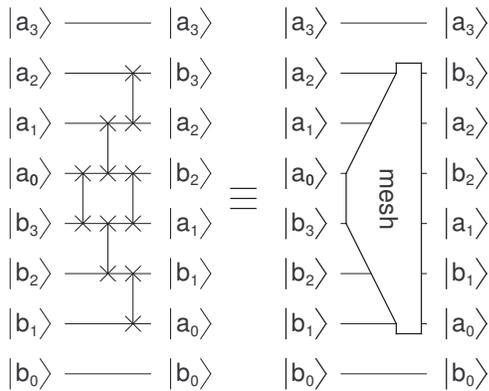}}
\end{center}
\caption{Circuit designed to interleave two quantum registers.}
\label{figure:mesh}
\end{figure}

Performing a controlled swap of two large registers is slightly
more difficult when only LNN interactions are available.  The two
registers need to be meshed so that pairs of equally significant
qubits can be controlled-swapped.  The mesh circuit is shown in
fig.~\ref{figure:mesh}.  This circuit element would not be
required in a general architecture.

\begin{figure}
\begin{center}
\resizebox{70mm}{!}{\includegraphics{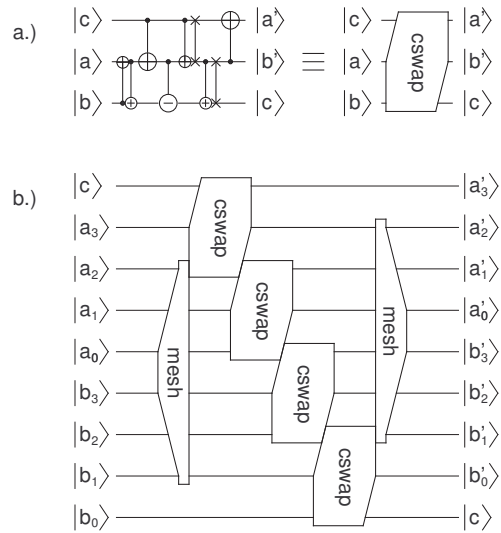}}
\end{center}
\caption{a.) LNN circuit for the controlled swapping of two qubits
$|a\rangle$ and $|b\rangle$.  The qubits $|a'\rangle$ and
$|b'\rangle$ represent the potentially swapped states.  b.) LNN
circuit for the controlled swapping of two quantum registers. Note
that when chained together, the effective depth of the cswap gate
is 4.} \label{figure:cswap}
\end{figure}

After the mesh circuit has been applied, the functional part of
the controlled swap circuit (fig.~\ref{figure:cswap}) can be
applied optimally with the control qubit moving from one end of
the meshed registers to the other.  The mesh circuit is then
applied in reverse to untangle the two registers.

The gate count and circuit depth of a mesh circuit is $L(L-1)/2$
and $L-1$ respectively.  The corresponding equations for a
complete LNN controlled swap are $L^{2}+5L$ and $6L$.  The general
controlled swap only requires $6L$ gates and can be implemented in
a circuit of depth $4L+2$.  The controlled swap is the only part
of this implementation of Shor's algorithm that is significantly
more difficult to implement on an LNN architecture.

\section{Modular Multiplication}
\label{mod_mult}

\begin{figure*}[p]
\begin{center}
\resizebox{!}{215mm}{\includegraphics{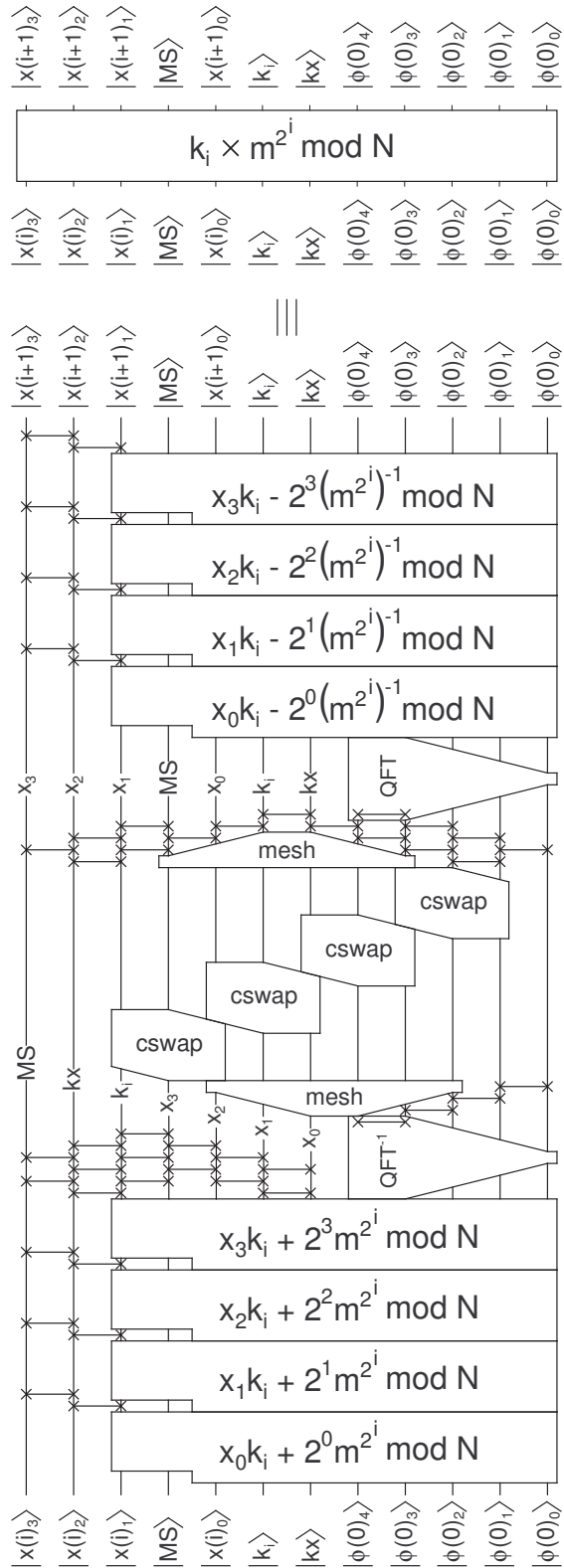}}
\end{center}
\caption{Circuit designed to modularly multiply $x(i)$ by
$m^{2^{i}}$ if and only if $k_{i}=1$.  Note that for simplicity
the circuit for $L=4$ has been shown.  Note that the bottom $L+1$
qubits are ancilla and as such start and end in the
$|\phi(0)\rangle$ state.  The swap gates within the two QFT
structures represent compound gates.} \label{figure:multiply}
\end{figure*}

The ideas behind the modular multiplication circuit of
fig.~\ref{figure:multiply} were discussed in section
\ref{shor_decomp}.  The first third comprises a controlled modular
multiply (via repeated addition) with the result being stored in a
temporary register. The middle third implements a controlled swap
of registers.  The final third resets the temporary register.

Note that the main way in which the performance of the LNN circuit
differs from the ideal general case is due to the inclusion of the
two mesh circuits.  Nearly all of the remaining swaps shown in the
circuit do not contribute to the overall depth.  Note that the two
swaps drawn within the QFT and inverse QFT are intended to
indicate the appending of a swap gate to the first and last
compound gates in these circuits respectively.

The total gate count for the LNN modular multiplication circuit is
$4L^{3}+20L^{2}+58L-2$ versus the general gate count of
$4L^{3}+13L^{2}+35L+4$. The LNN depth is $16L^{2}+40L-7$ and the
general depth $16L^{2}+33L-6$.

\section{Complete Circuit}
\label{comp_circ}

\begin{figure*}[p]
\begin{center}
\resizebox{!}{215mm}{\includegraphics{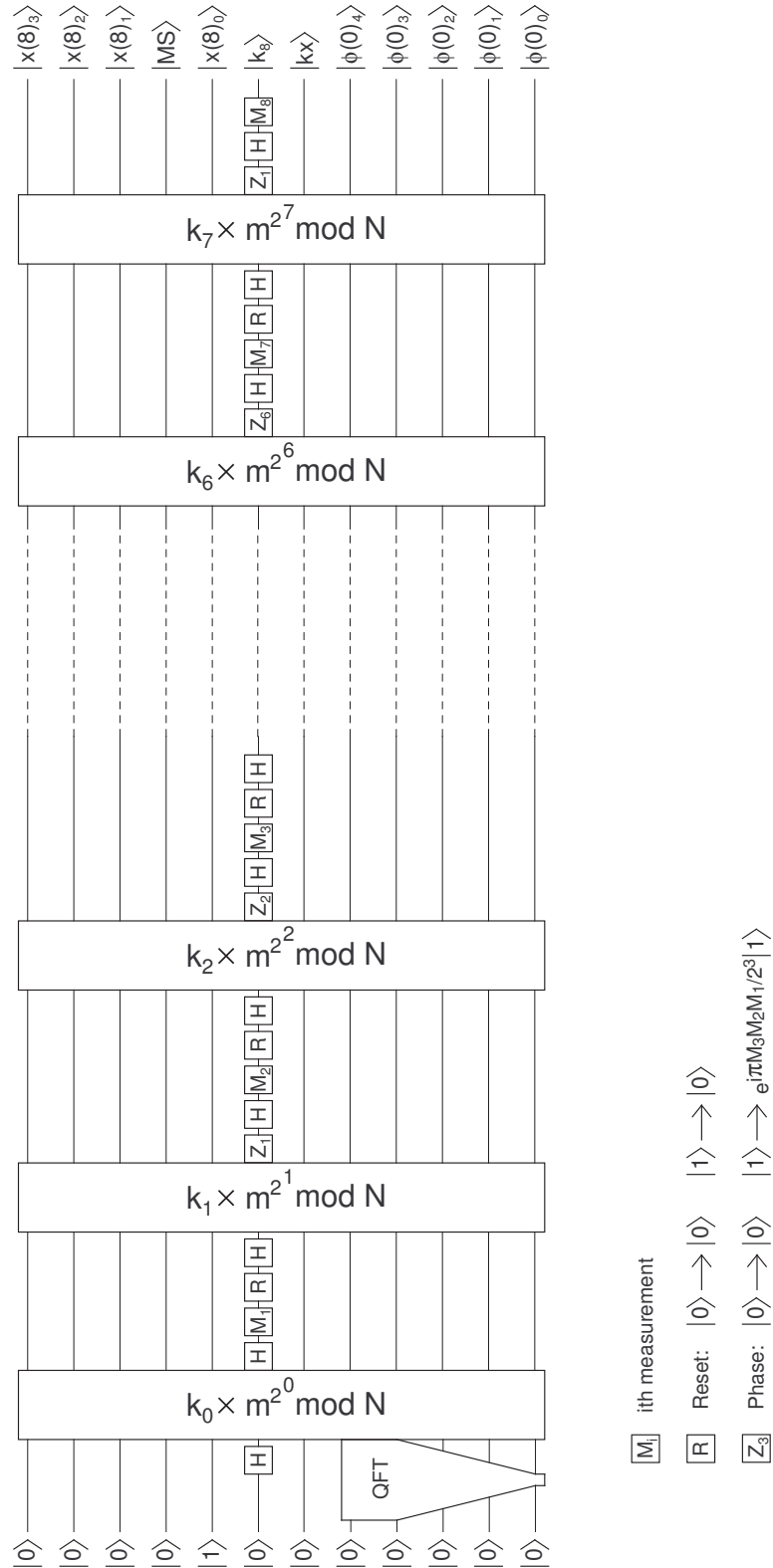}}
\end{center}
\caption{Circuit designed to compute Shor's algorithm.  The
single-qubit gates interleaved between the modular multiplications
comprise a QFT that has been decomposed by using measurement gates
to remove the need for controlled quantum phase rotations.  Note
that without these single-qubit gates the remaining circuit is
simply modular exponentiation.} \label{figure:modexp}
\end{figure*}

The complete circuit for Shor's algorithm
(fig.~\ref{figure:modexp}) can best be understood with reference
to fig.~\ref{figure:serial_parallel_qft}a and the five steps
described in section \ref{shor_alg}.  The last two steps of Shor's
algorithm are a QFT and measurement of the qubits involved in the
QFT.  When a 2-qubit controlled quantum gate is followed by
measurement of the controlled qubit, it is equivalent to measure
the control qubit first and then apply a classically controlled
gate to the target qubit.  If this is done to every qubit in
fig.~\ref{figure:serial_parallel_qft}a, it can be seen that every
qubit is decoupled.  Furthermore, since the QFT is applied to the
$k$ register and the $k$ register qubits are never interacted with
one another, it is possible to arrange the circuit such that each
qubit in the $k$ register is sequentially used to control a
modular multiplication, QFTed, then measured.  Even better, after
the first quit of the $k$ register if manipulated in this manner,
it can be reset and used as the second qubit of the $k$ register.
This one qubit trick \cite{Park00} forms the basis of
fig.~\ref{figure:modexp}.

The total number of gates required in the LNN and general cases
are $8L^{4}+40L^{3}+116\frac{1}{2}L^{2}+4\frac{1}{2}L-2$ and
$8L^{4}+26L^{3}+70\frac{1}{2}L^{2}+8\frac{1}{2}L-1$ respectively.
The circuit depths are $32L^{3}+80L^{2}-4L-2$ and
$32L^{3}+66L^{2}-2L-1$ respectively.  The primary result of this
paper is that the gate count and depth equations for both
architectures are identical to first order.

\section{Conclusion}
\label{conc}

We have presented a circuit implementing Shor's algorithm in a
manner appropriate for a linear nearest neighbour qubit array and
studied the number of extra gates and consequent increase in
circuit depth such a design entails.  To first order our circuit
involves $8L^{4}$ gates arranged in a circuit of depth $32L^{3}$
on $2L+4$ qubits --- figures identical to that possible when
interactions between arbitrary pairs of qubits are allowed.  Given
the importance of Shor's algorithm, this result supports the
widespread experimental study of linear nearest neighbour
architectures.

Simulations of the robustness of the circuit when subjected to
noise are in progress.  Future simulations will investigate the
performance of the circuit when protected by LNN quantum error
correction.

\bibliography{../References} 

\widetext

\end{document}